\documentstyle[epsfig,epsf]{article}

\def\nmnt{\nu_\mu \longleftrightarrow \nu_\tau}
\def\nmns{\nu_\mu \longleftrightarrow \nu_{sterile}}
\def\nmne{\nu_\mu \longleftrightarrow \nu_e}

\def\ne{\nu_e}
\def\nm{\nu_\mu}
\def\nt{\nu_\tau}
\def\s2t{\sin^2 2 \theta}
\def\Dm2{\Delta m^2}

\begin{document}

\large

\begin{center}
{\Large {\bf Atmospheric neutrino oscillations with the MACRO detector}}
\end{center}

\vskip .5 cm

\begin{center}
M. Giorgini, for the MACRO Collaboration{\footnote{For the 
full list of the Coll., see ref. \cite{sterile}}}

 \vspace{2mm}

{\normalsize {\it University of Bologna and INFN, 
 Viale C. Berti Pichat 6/2 \\ I-40127 Bologna, Italy} \\ 
E-mail: miriam.giorgini@bo.infn.it}

\par~\par
  
{\normalsize Talk given at Beyond the Desert 02, Conference on Physics beyond 
the Standard Model, \\ Oulu, Finland, 2-7 June 2002}

\vskip .7 cm
{\normalsize \bf Abstract}
\end{center}

{\normalsize After a short overview of the MACRO detector, located at the  
Gran Sasso Laboratory, I shall present and discuss the results on 
atmospheric muon neutrino oscillations, concerning low ($\langle 
E_\nu \rangle \sim 4$ GeV) and 
high ($\langle E_\nu \rangle \sim 50$ GeV) energy data. Using the
Multiple Coulomb Scattering of muons inside the lower part of the detector,
  estimates of the neutrino energy were made event by event for the high 
energy sample. The
 data on angular distributions, absolute flux and $L/E_\nu$
distribution favor $\nmnt$ oscillations with
maximal mixing and $\Dm2=2.5 \cdot 10^{-3}$ eV$^2$.}

\section{Introduction}
\label{sec:introduction}

The phenomenon of neutrino oscillations \cite{pontecorvo} implies that 
the flavour states 
 $\nu_l$ are linear combinations of the mass eigenstates $\nu_m$ 
through the elements of the unitary mixing matrix $U_{lm}$:
\begin{equation}
\nu_l = \sum_{m=1}^3 U_{lm}\ \nu_m
\end{equation}

 In the simple case of two-flavour oscillations $(\nm,\nt)$, the 
oscillation probability of a $\nm$ into a $\nt$ can be expressed by:
\begin{equation}
P(\nm \to \nt)=\s2t \cdot \sin^2 \left( \frac{1.27 \cdot \Dm2 L}{E_\nu} \right)
\label{eq:p_osc}
\end{equation}
where $\Dm2=m_3^2 -m_2^2$ (eV$^2$), $\theta$ is the mixing angle, $L$ (km) 
is the traveled neutrino distance and $E_\nu$ (GeV) is the neutrino energy.

\begin{figure}
\begin{center}
\scalebox{0.6}{
\includegraphics{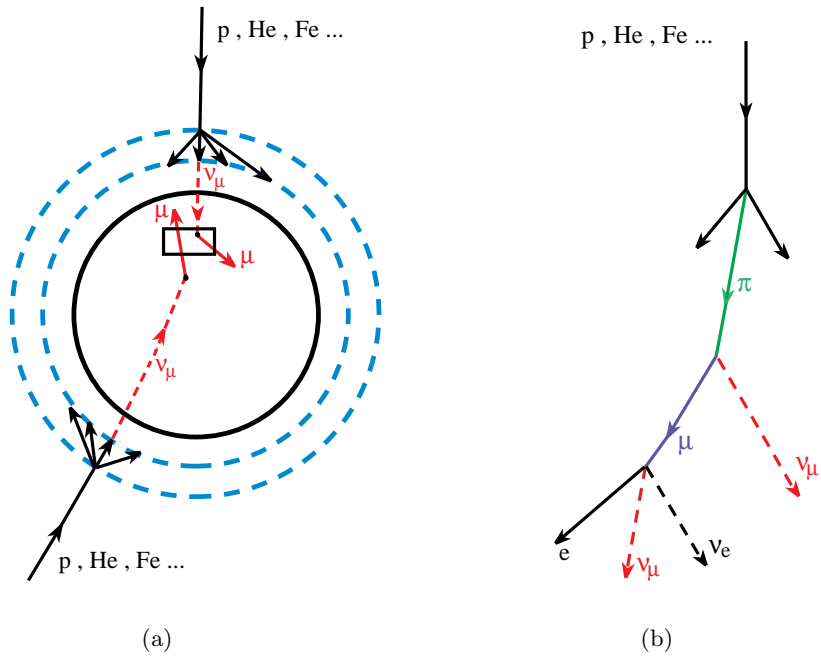}}
\end{center}
{\small \hspace{4.5cm} (a) \hspace{6cm} (b)}
\caption{(a) Illustration of the production, travel and interactions
of atmospheric muon neutrinos; (b) interaction of a primary
cosmic ray in the upper atmosphere, production of pions (and kaons) and 
their decays leading to the atmospheric $\ne$ and $\nm$.}
\label{fig:catena}
\end{figure}

Atmospheric neutrinos are the result of a decay chain 
starting with the interactions of high energy primary cosmic rays in the 
upper atmosphere. Each interaction 
produces a large number of pions and kaons, which decay yielding  
muons and muon neutrinos; also the muons decay yielding muon and electron 
neutrinos. These neutrinos are produced at about 20 km above ground and 
they proceed towards the Earth. An underground detector is ``crossed" by
a neutrino flux from all directions. Fig.~\ref{fig:catena}a shows how an 
underground detector can detect and study atmospheric muon 
neutrinos and Fig.~\ref{fig:catena}b is a simple illustration of the 
described decay chain.

Atmospheric neutrinos are well suited to study
neutrino oscillations, since they have energies from a fraction of a GeV up 
to more than 100 GeV and they may travel distances $L$ from few tens of km 
up to 13000 km; thus the $L/E_\nu$ values range from $\sim 1$ km/GeV to more 
than $10^4$ km/GeV. 

 In the following, I shall briefly describe the main features of the 
MACRO detector, then I shall discuss
the main results of MACRO about muon neutrino oscillations.

\section{The MACRO experiment}
\label{sec:MACRO}
MACRO \cite{technical} was a large multipurpose detector which was 
operational at the Gran 
Sasso Laboratory from 1989 till the end of 2000. Though it was optimized 
to search for the supermassive GUT magnetic monopoles, the MACRO research 
program included a wide range of topics, like atmospheric $\nm$ 
oscillations, astrophysics, nuclear and particle physics and cosmic ray 
physics. 

\begin{figure}
\begin{center}
\scalebox{0.65}{
\includegraphics{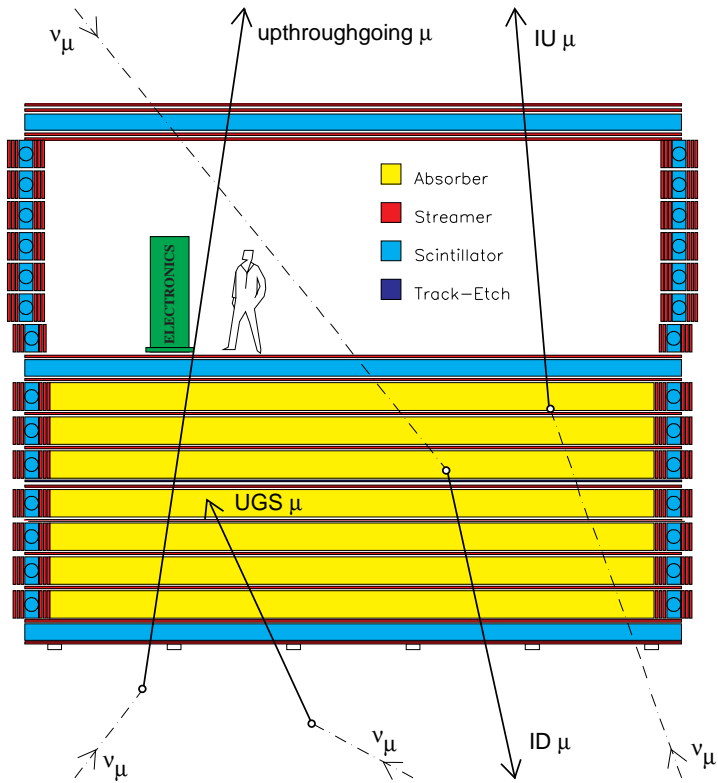}}
\end{center}
\caption{Vertical section of the MACRO detector. Event topologies induced
by $\nm$ interactions in or around the detector: IU = semicontained 
Internal Upgoing $\mu$; ID = Internal Downgoing $\mu$; UGS = Upgoing 
Stopping $\mu$; Upthroughgoing = upward throughgoing $\mu$.}
\label{fig:macro}
\end{figure}

The average rock overburden of 3700 hg/cm$^2$ reduced the atmospheric muon 
flux by a factor $\sim 10^6$.
MACRO was a large rectangular box, 76.6m$\times$12m$\times$9.3m, divided
longitudinally in six supermodules and vertically in a lower
part (4.8 m high) and an upper part (4.5 m high), Fig.~\ref{fig:macro}.
 It had three types of detectors which gave redundancy of informations: liquid
 scintillation counters, limited streamer tubes and nuclear track 
detectors. This last detector was used only for rare particle searches. 

For muon and for neutrino physics and astrophysics studies, 
 the streamer tubes were used for muon tracking 
and the liquid scintillation counters for fast timing. The lower part of
the detector was filled with trays of crushed rock absorbers alternating 
with streamer tube planes; the upper part was open and contained the 
electronics \cite{technical}. 

There were 10 horizontal planes of streamer tubes in the bottom half of the 
detector, and 4 planes at the
top, all with wires and 26.5$^\circ$ stereo strips readout. Six vertical 
planes of streamer tubes and one layer of scintillators covered each 
side of the detector.
 The scintillator system consisted of three layers 
of horizontal counters, and of the mentioned  vertical layer
along the sides of the detector. 

 The combination of the informations from the 
streamer tubes and from the scintillators allowed tracking with a precision 
of 1 cm over path lengths of several meters, and timing with a precision of 
600 ps. The detector provided a total acceptance $S \Omega \simeq 10000$ 
m$^2$ sr for an isotropic flux of particles. 

 Fig.~\ref{fig:macro} is a vertical section of the detector; it shows 
a general view of the detector and gives a sketch of the different 
topologies of detected neutrino-induced muon
events used to study neutrino oscillations: Upthroughgoing muons, Internal
Upgoing muons (IU), Upgoing Stopping muons (UGS) and Internal 
Downgoing muons (ID).

The {\it Upthroughgoing muons} \cite{high} (with $E_\mu > 1$ GeV) come from  
interactions in the rock below the detector of muon neutrinos with an average
energy $\langle E_\nu \rangle \sim 50$ GeV. The tracking is performed 
with streamer tubes 
hits; the time information, provided by scintillation counters,
 allows the determination of the direction (versus) by the time-of-flight 
(T.o.F.) method. 

The {\it semicontained upgoing muons} (IU) \cite{low} come from
$\nm$ interactions inside the lower apparatus. Since two
scintillation counters are intercepted, the T.o.F. method
is applied to identify the upward going muons. The average 
parent neutrino energy for these events is $\sim 4$ GeV. 
If the atmospheric neutrino anomalies are the results of
$\nm$ oscillations with maximal mixing
and $10^{-3} < \Dm2 < 10^{-2}$ eV$^2$, one expects a 
reduction of about a factor two in the flux of IU events, 
without any distortion in the shape of the angular distribution.

The {\it up stopping muons} (UGS) \cite{low} are due to $\nm$ interactions 
in the rock below the detector yielding upgoing muon tracks stopping in the 
detector; the 
{\it semicontained downgoing muons} (ID) \cite{low} are due to 
$\nm$ induced downgoing tracks with vertex in the lower MACRO.
The events are found by means of topological criteria; the lack
of time information prevents to distinguish between the two subsamples.
An almost equal number of UGS and ID events is expected in the no 
oscillation hypothesis. In case of oscillations,
 a  similar reduction in the flux of the up stopping events 
and of the semicontained upgoing muons is expected; no reduction is 
instead expected for the semicontained downgoing events (which come from
neutrinos which traveled $\sim 20$ km).

\section{Upthroughgoing muons}
\label{sec:upthroughgoing}

The data were collected during the running period from  March 1989 
to April 1994 with the detector under construction and during the runs with
the complete 
detector from 1994 until December 2000 (livetime 5.52 yrs). Since the 
total livetime normalized to the full configuration is 6.16 yrs, the
statistics is largely dominated by the full detector run. The analysis of
a data sample of more than 40 million atmospheric downgoing muons
 achieved a rejection factor of $\sim 10^{-7}$ which includes background 
caused by showering events and radioactivity in coincidence with muons.

One of the main cuts to remove background requires that the position 
of a muon hit in
each scintillator, as determined from the timing within the scintillator
counter, agrees within $\pm 70$ cm with the position indicated by the 
streamer tube track. 

Downgoing muons passing near MACRO may produce low energy 
upgoing particles, which could appear neutrino-induced upgoing muons 
\cite{background}. In order to reduce this background, we require that each
upgoing muon crosses at least 200 g cm$^{-2}$ of material in the lower 
part of the detector. 

\begin{figure}
\begin{center}
\scalebox{0.8}{
\includegraphics{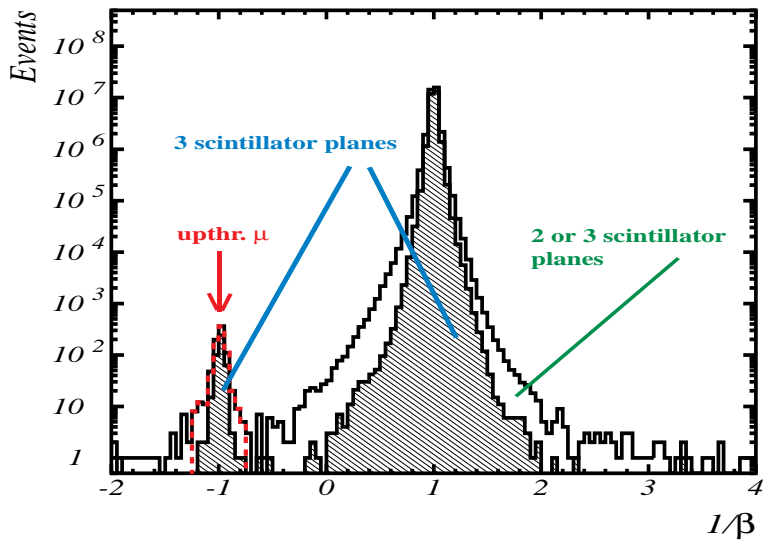}}
\end{center}
\caption{$1/\beta$ distributions for throughgoing (downgoing and upgoing)
 muons collected with the complete detector. The shaded 
areas concern muons crossing 3 scintillation counters, while the open
histogram concerns muons crossing 2 or 3 counters. Two vertical dotted lines
delimit the range $-1.25 \leq 1/\beta \leq -0.75$ of upthroughgoing 
muons. There are $\sim 3.5\cdot 10^7$ downgoing muons with $1/\beta \sim 1$ 
and 863 (before subtraction of background) upgoing muons with 
$1/\beta \sim -1$.}
\label{fig:1_su_beta}
\end{figure}

A large number of nearly horizontal $(\cos \Theta > -0.1)$ upgoing muons
have been observed coming from azimuth angles between $-30^\circ$ and 
$120^\circ$. In this direction, the rock overburden is insufficient to
remove nearly horizontal downgoing muons which have scattered in the mountain 
and appear as upgoing. This region was excluded for real events and for
MC simulations. 

For muons crossing 3 scintillation counters a linear fit of the times 
as a function of the path length is performed and a cut is applied on the 
$\chi^2$; further minor cuts are applied to events crossing 2 counters. 

The direction of muons crossing MACRO is determined by the time of flight
method, between two layers of scintillators. The measured muon velocity is
calculated with the convention that downgoing muons have $1/\beta=+1$ and
upgoing muons have $1/\beta=-1$. The $1/\beta$ distribution for the sample 
collected with the full detector is shown in Fig.~\ref{fig:1_su_beta}. We 
selected upwardgoing muons requiring $-1.25 \leq 1/\beta \leq -0.75$; we 
found 863 events, of which 809 events remained after quality cuts and 
background subtraction. 

In the upgoing muon simulation 
the neutrino flux computed by the Bartol group \cite{Agrawal96} was used.
The cross sections for the neutrino interactions was calculated
using the deep inelastic parton distributions  \cite{Gluck95}.
 The muon propagation to the detector was computed using the
energy loss calculation in standard rock \cite{Lohmann85}.

The total systematic uncertainty on the
expected flux of upthroughgoing muons, arising from the neutrino flux, cross 
section and muon propagation, was estimated to be $\sim$17\%. 
This theoretical uncertainty is mainly a scale error that does not affect 
the shape of the angular distribution; the error on the shape is 
$\sim 5\%$. The same cuts applied to the data 
were used for the simulated events: assuming no oscillations, they selected 
1122 MC events. 

\begin{figure}
\begin{center}
\scalebox{0.9}{
\includegraphics{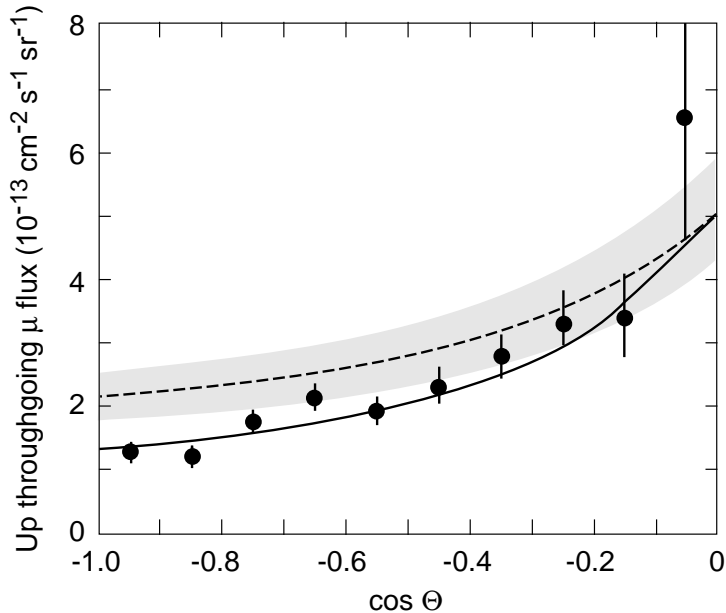}}
\end{center}
\caption{Zenith distribution of the upthroughgoing muons (809 
events, background subtracted). The data (black points) have statistical 
and systematic errors added in quadrature. The shaded region indicates the 
theoretical scale error band of $\pm 17\%$ (see text) associated to the
no oscillation prediction (dashed line). The solid line is the 
fit to an oscillated flux with maximal mixing and $\Dm2 = 2.5\cdot 10^{-3}$ 
eV$^2$.} 
\label{fig:cos-uptr}
\end{figure}

 Fig.~\ref{fig:cos-uptr} shows the zenith angle $(\Theta$) distribution 
of the measured flux of upthroughgoing muons (black points); the MC 
expectation for no oscillations is indicated by the dashed line with the
17\% scale error band. A deficit of events, more evident around the vertical 
direction, can be noticed. The ratio of the observed number of events to 
the expectation without oscillations in $-1 < \cos \Theta < 0$ is
$R_{upthr}=0.721 \pm 0.026_{stat} \pm 0.043_{sys} \pm 0.123_{th}$.

The shape of the angular distribution was tested with the hypothesis of no 
oscillations normalizing
the prediction to the data. The $\chi^2/D.o.F.$ corresponds to a probability
of 0.2\%. In the hypothesis of two-flavour ($\nmnt$)
oscillations, the minimum $\chi^2/D.o.F.$ in the physical region is 
9.7/9 ($P=37\%$)
for maximal mixing and $\Dm2 = 2.5\cdot 10^{-3}$ eV$^2$. An
independent test was made on the number of the events. 

 Combining the probabilities from the two independent tests, the best
probability is 66\% for maximal mixing and $\Dm2 \simeq 2.4\cdot 10^{-3}$ 
eV$^2$. The result of the fit is the solid line in Fig.~\ref{fig:cos-uptr}.

\subsection{Two-flavour {\boldmath $\nmns$} oscillations}
\label{sec:sterile}
The simple relation (\ref{eq:p_osc}) should be modified when neutrinos
propagate through matter \cite{matter}. The weak potential in matter 
produces a phase 
shift that could modify the oscillation pattern if the oscillating
neutrinos have different interactions with matter. The matter effect could
discriminate between different neutrino channels: it could be
important for $\nmne$ and for $\nmns$ oscillations, while for $\nmnt$
there is no matter effect ($\nm$ and $\nt$ have the same weak potential 
in matter). Then, it is important when $E_\nu/|\Dm2| \geq
10^3$ GeV/eV$^2$, therefore for high energy events \cite{matter}.

\begin{figure}
\begin{center}
\scalebox{0.83}{
\includegraphics{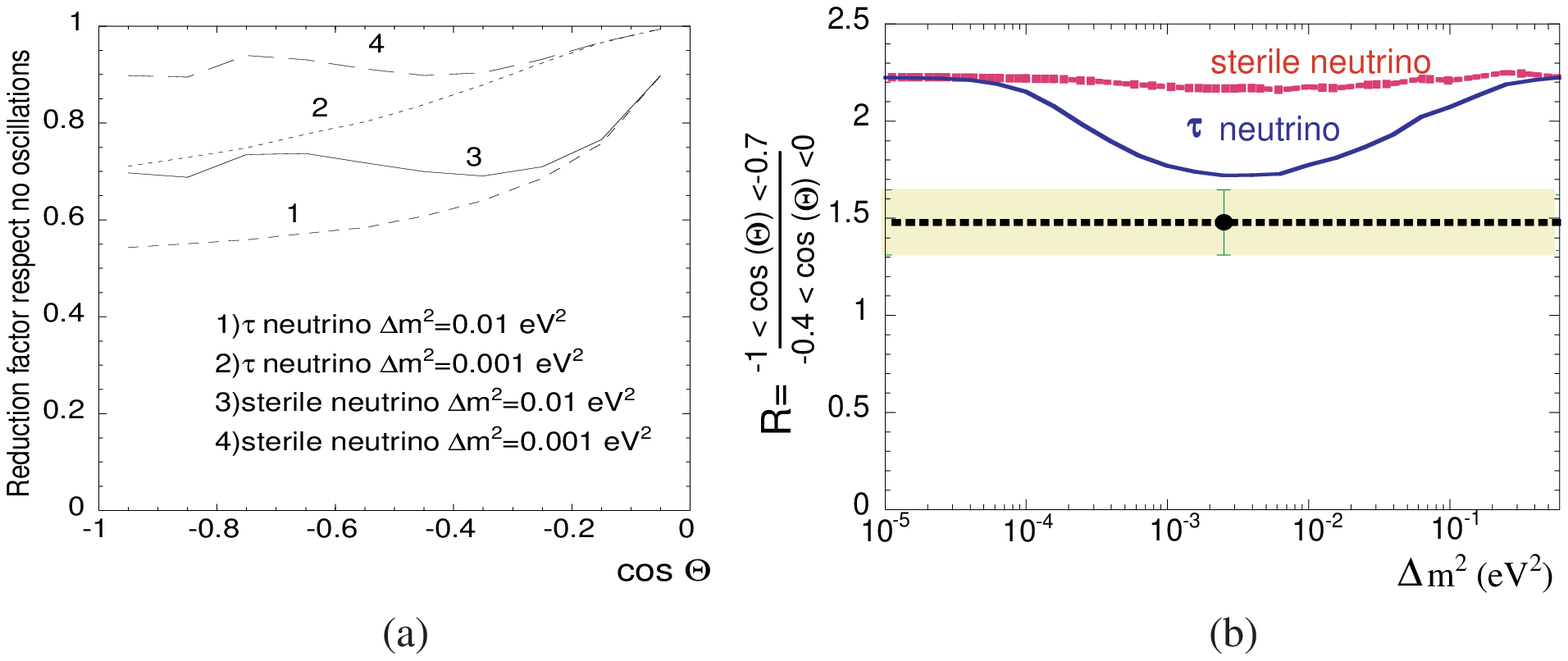}}
\end{center}
\caption{(a) Reduction factor with respect to no oscillations as a function of
the zenith angle $\Theta$, computed for $\nmnt$ and $\nmns$ oscillations 
with maximal mixing and $\Dm2=10^{-3}$ and $\Dm2=10^{-2}$. (b)  Ratio of 
MACRO events with $-1 < \cos \Theta < -0.7$ to events with
 $-0.4 < \cos \Theta < 0$ as a function of $\Dm2$ for maximal mixing.
 The black point with error bar is the measured value, the predictions 
for  $\nmnt$ and $\nmns$ oscillations are also shown.}
\label{fig:sterileneutrino}
\end{figure}

In the $\nmns$ oscillation scenario, the matter effect changes the shape
 of the angular distribution and the total number of events with respect
to the vacuum oscillations. Large matter effects are expected for 
neutrinos crossing the Earth, due to the long path length and to the 
increase of the density in the Earth core. In absence of resonances, due to 
particular values of the oscillation parameters, the matter effect produces
a reduction of the oscillation effect, giving a prediction similar to the 
no oscillation scenario, particularly for directions near the vertical.     
 Fig.~\ref{fig:sterileneutrino}a shows the predicted reduction for
$\nmnt$ and $\nmns$ oscillations with maximal mixing and $\Dm2=10^{-3}$ 
eV$^2$ or $\Dm2=10^{-2}$ eV$^2$. 

A statistically powerful test is based on the ratio between the events 
with $-1 < \cos \Theta < -0.7$ and the events with $-0.4 < \cos \Theta < 0$
\cite{sterile}.   
 This quantity is more powerful than the $\chi^2$ because the data are
binned to maximize the difference between the two hypotheses and the
ratio is sensitive to the sign of the variation. The angular regions were
chosen by MonteCarlo (MC) methods in order to have the best discrimination 
between the $\nmnt$ and $\nmns$ oscillations.

 In this ratio, most of the theoretical uncertainties
on neutrino flux and cross sections cancel. The remaining theoretical 
error was estimated at $\leq 5\%$; the systematic experimental error on 
the ratio, due to analysis
cuts and detector efficiencies, is 4.6\%. Combining the experimental
and theoretical errors in quadrature, a global estimate of 7\% is obtained.
MACRO measured 305 events with $-1 < \cos \Theta < -0.7$ and 206 with
$-0.4 < \cos \Theta <0$; the ratio is $R=1.48 \pm 0.13_{stat} \pm 0.10_{sys}$
\cite{sterile}.
 The result is shown in Fig.~\ref{fig:sterileneutrino}b as the black 
point, with the error bar including statistical and systematic errors;
 the predictions for $\nmnt$ and $\nmns$ oscillations as a function 
of $\Dm2$ are also shown.

For $\Dm2=2.5\cdot 10^{-3}$ eV$^2$ and maximal mixing, 
 the minimum expected value of the ratio for $\nmnt$ oscillations
is $R_\tau=1.72$; for $\nmns$ oscillations one expects
$R_{sterile}=2.16$. The maximum probabilities $P_{best}$
to find a value of $R_\tau$ and of $R_{sterile}$ smaller than $R_{expected}$ 
are 9.4\% and 0.06\% respectively. Hence, the ratio of the maximum 
probabilities is $P_{best_\tau}/P_{best_{sterile}}=157$, so that 
$\nmns$ oscillations (with any mixing) are excluded at 99\% C.L.
compared to the $\nmnt$ channel with maximal mixing and 
$\Dm2=2.5\cdot 10^{-3}$ eV$^2$ \cite{sterile}.

\subsection{{\boldmath $\nm$} energy estimates by Multiple Coulomb 
Scattering of upthroughgoing muons} 
\label{sec:mcs}

Since MACRO was not equipped with a magnet, the only way to experimentally 
estimate the muon energy was through the Multiple Coulomb Scattering (MCS) 
of muons in the absorbers, see Fig.~\ref{fig:macro}. The r.m.s. of the 
lateral displacement of a relativistic muon  traveling for a distance 
$X$ can be written as:

\begin{equation}
\sigma_{MCS} \simeq \frac{X} {\sqrt{3}} \frac{13.6 \cdot 10^{-3} GeV}
{p\beta c} \sqrt{X/X^0} \cdot (1+0.038 \ln (X/X^0))
\end{equation}
where $p$ (GeV/c) is the muon momentum  and $X/X^0$ is the amount of crossed
material in units of radiation lengths. A muon crossing the whole
apparatus on the vertical has $\sigma_{MCS} \simeq 10 $(cm)/$E_\mu$(GeV). The 
muon energy estimate can be performed up to a saturation point, occurring when
$\sigma_{MCS}$ is comparable with the detector space resolution. 

Two analyses were performed. 

The  first analysis was made studying the deflection of upthroughgoing muons 
in MACRO with the streamer tubes in digital mode \cite{mcs-spurio}. Using 
MC methods to estimate
the muon energy from its scattering angle, the data were divided into 3
subsamples with different average energies, in 2 samples in zenith angle 
$\Theta$ and finally in 5 subsamples with different
average values of $L/E_\nu$. This method could reach a spatial resolution of 
$\sim 1$ cm, which implies a saturation point at $E_\mu \simeq 10$ GeV.

\begin{figure}
\begin{center}
\scalebox{0.7}{\includegraphics{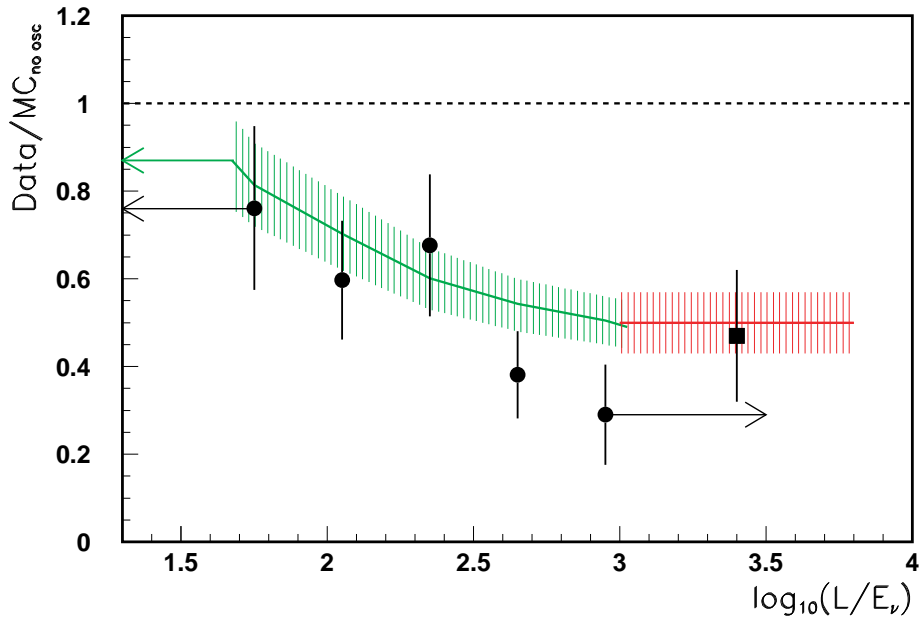}}
\end{center}
\caption {Ratio Data/MC$_{\mbox {no~osc}}$ as a function of the   
$\log_{10} (L/E_\nu)$ estimated using the streamer tubes in ``drift mode".
 The upthroughgoing data are the points with error bars, the continuous line
and the shaded region (12\% point-to-point systematic error) represent the 
MC predictions for $\nmnt$ oscillations, the black square refers to IU 
events.} 
\label{fig:mcs}
\end{figure}

As the interesting energy region for atmospheric neutrino 
oscillations spans from $\sim 1 $ GeV to some tens of GeV, it was important
to improve the spatial resolution of the detector to push the saturation
point as high as possible. For this purpose, a second analysis was performed
using the streamer tubes in ``drift mode" \cite{mcs-scap} using the TDC's 
included in the QTP system \cite{qtp}, originally designed for the search for 
magnetic monopoles. To check the electronics and the feasibility of the
analysis, two ``test beams'' were performed at the CERN PS-T9 and SPS-X7
beams. For each triggered tube, the arrival time
of the signal multiplied by the drift velocity, measured at the test beam 
\cite{mcs-miriam}, gave the radius of the drift circle. The
 upthroughgoing muon tracks were reconstructed as the best fit of the
drift circles referring to the triggered streamer tubes. The space resolution
achieved is $\simeq 3$ mm, a factor of 3.5 better than in the 
previous analysis. 
 For each muon, 7 MCS sensitive variables were given in input to a Neural 
Network (NN) which was previously trained to estimate $E_\mu$ with MC events 
of known input energy crossing the detector at different zenith angles. 

The calibration procedure of the NN output used to reconstruct event
by event $E_\mu$ was used also to estimate the corresponding value of 
$E_\nu$. The collected informations allowed to 
separate the whole sample of upthroughgoing muons in 4 subsamples with 
average energies $E_\nu$ of 12, 20, 50 and 102  
GeV, respectively. The comparison of the
zenith angle distributions of the 4 energy subsamples with the 
predictions of no oscillation MC shows a
strong disagreement at low energies (where there is a deficit of vertical 
events), while the agreement is restored at higher neutrino energies. 

For each event, the neutrino traveled distance $L$ was measured with 
a $\sim 3\%$ 
precision by using the reconstructed zenith angle of the tracked muon.  
 The distribution of the ratios $R=($Data/MC$_{\mbox{no~osc}})$ 
obtained by the second method is plotted in Fig.~\ref{fig:mcs}
as a function of $\log_{10} (L/E_\nu)$. The black points are the data, their 
error bars contain the statistical error, the 17\% uncertainty on the 
predicted 
flux and a 12\% point-to-point systematic error added in quadrature. The 
continuous line represents the MC prediction for $\nmnt$ oscillations 
with maximal mixing and $\Dm2 =2.5\cdot 10^{-3}$ eV$^2$, the shaded area 
includes the 12\% point-to-point error associated to the MC prediction. 
 In the same figure, the black square and the associated MC prediction
refer to the Internal Up events (see Sect. \ref{sec:lowenergy}). 

\section{Low energy events}
\label{sec:lowenergy}
These events were mainly due to $\nm$ CC interactions, with a contribution 
from NC and $\ne$ ($\sim 13\%$ for IU and $\sim 10\%$ for UGS+ID). The data 
concern only the running period with the detector
in the full configuration from April 1994 to December 2000. 
 Due to the difference between 
the topologies of the low energy events, two separate analyses were performed.

The identification of Internal Upgoing (IU)  events was based both on 
topological criteria and T.o.F. measurements. The IU sample corresponds to 
an effective livetime of 5.8 yrs. The basic request was the presence of a 
streamer tube track reconstructed in space matching at least two hits in two
different scintillators in the upper part of the apparatus. The track 
starting point had to be inside the apparatus. To reject fake semicontained 
events entering from a detector crack, the extrapolation of the track in 
the lower part of the detector had to cross and not fire at least three
streamer tube planes and one scintillation counter. The measured muon 
velocity $\beta c$ was evaluated with the same convention 
of upgoing muons. The range of the IU signal is 
$-1.3 \leq 1/\beta \leq -0.7$. After the subtraction of background 
events, mainly due to wrong time measurements or secondary particles hits, we
had 154 upgoing partially contained events. 

\begin{figure}
\begin{center}
\scalebox{0.85}{
\includegraphics{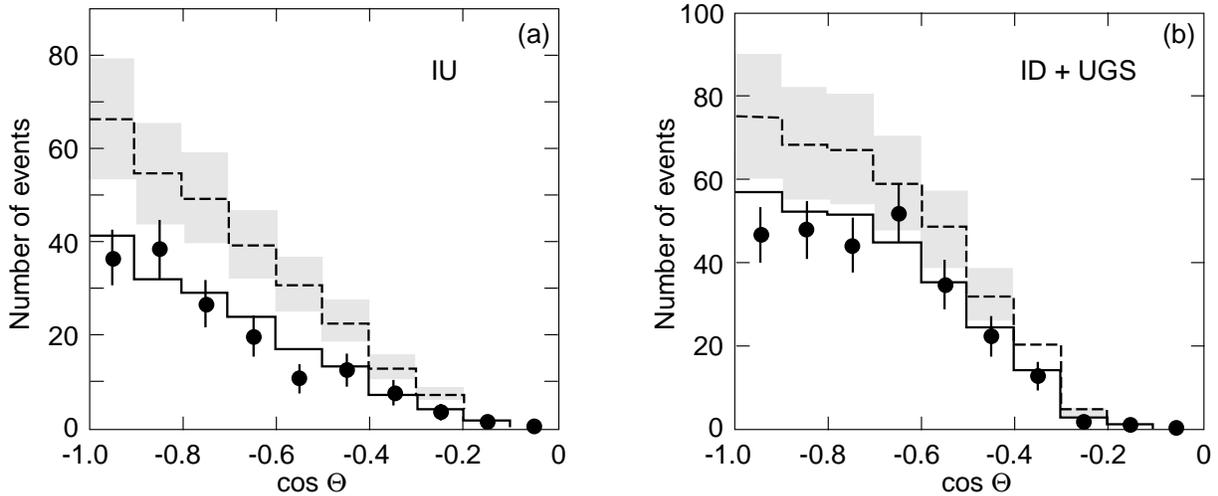}}
\end{center}
\caption {Measured zenith distributions (a) for the IU events 
 and (b) for the ID+UGD events. The black points are the data, the dashed 
lines and the shaded regions (25\% uncertainty)
correspond to MC predictions assuming no oscillations. The full line is 
the expectation for $\nmnt$ oscillations with 
maximal mixing and $\Dm2 =2.5\cdot 10^{-3}$ eV$^2$.}
\label{fig:low_cosze}
\end{figure}

The identification of ID+UGS events was based on topological criteria. The
candidates had a track starting (ending) in the lower apparatus and crossing
the bottom detector face. The track had also to be located or oriented in such
a way that it could not have entered (exited) undetected through 
insensitive regions of the apparatus. For this analysis the effective livetime 
was 5.6 yrs. The event selection required the presence of one reconstructed
track crossing the bottom layer of the scintillators and that all hits along
the track were confined one meter inside each MACRO supermodule. To reject
ambiguous or wrongly tracked events passing the selection, a scan with the
Event Display was performed. After background 
subtraction, we had 262 ID+UGS events.
 
The MC simulations for the low energy data used the Bartol neutrino
flux \cite{Agrawal96} and the low energy neutrino cross sections 
\cite{lipari94}. We estimated a total theoretical scale uncertainty on the 
predicted number of muons of the order of 25\% (probably overestimated). 

With the full MC simulation, the prediction for IU events was 
$285 \pm 28_{syst} \pm 71_{th}$, while the observed number of events was
$154 \pm 12_{stat}$. The ratio was 
$R_{IU}=$ (Data/MC)$_{IU} = 0.54 \pm 0.04_{stat} \pm 0.05_{syst} 
\pm 0.13_{th}$. For this sample, the average value of $\log_{10}(L/E_\nu)$ is
3.2, in good agreement with the hypothesis of neutrino oscillations. The black
square in Fig. \ref{fig:mcs} represents this value.

 The prediction for ID+UGS events was $375 \pm 37_{syst} \pm 94_{th}$, while
the observed number of events was $262 \pm 16_{stat}$. The ratio was 
$R_{ID+UGS}=$  (Data/MC)$_{ID+UGS}=0.70 \pm 0.04_{stat} \pm 0.07_{syst} 
\pm 0.17_{th}$. 

\begin{figure}
\begin{center}
\scalebox{0.6}{\includegraphics{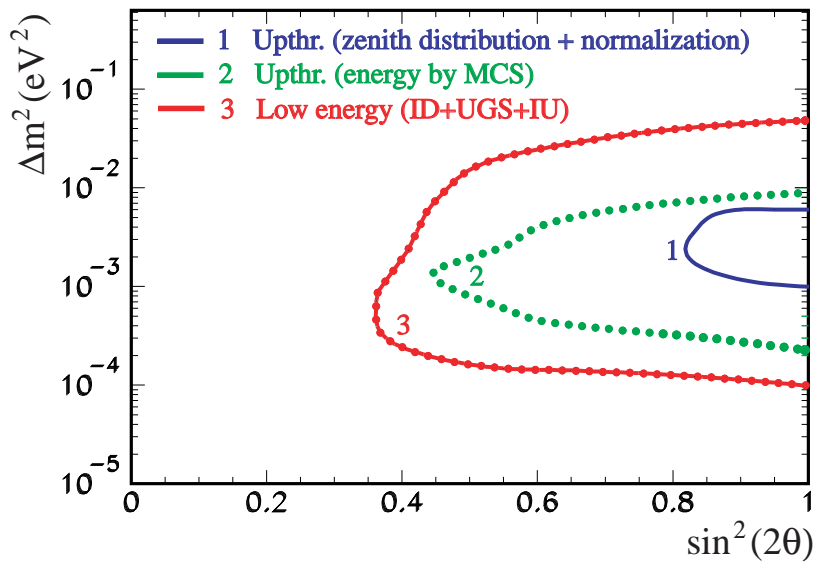}}
\end{center}
\caption {Allowed regions in the $\s2t - \Dm2$ plane computed using the 
Feldman-Cousins \cite{feldman-cousins} procedure. The curves 1 and 2
refer to upthroughgoing muons using the angular distribution + 
normalization and the muon energy estimate via MCS, respectively. The curve
3 refers to low energy events.} 
\label{fig:contour}
\end{figure}

 The angular distributions of low energy events are
shown in Fig. \ref{fig:low_cosze}. The measured data (black points) are in 
good agreement with the predicted angular distributions based on $\nmnt$ 
oscillations with the parameters obtained from the upthroughgoing muon
sample (full histogram in Fig.~\ref{fig:low_cosze}).

\section{Exclusion plots in the {\boldmath $\s2t - \Dm2$} plane}
\label{sec:exclusion}
For each event category, the 90\% C.L. regions were computed using the 
Feldman-Cousins procedure \cite{feldman-cousins}, Fig. \ref{fig:contour}.
 For upthroughgoing muons, we obtained the curve 1 using the angular
distribution and the normalization and the curve 2 using only the muon energy
reconstructed with the MCS method (Sect. \ref{sec:mcs}). For low energy events
 we used a $\chi^2$ obtained from the shape of the angular distribution, the
total normalization and the ratio between the two topologies of events. The
low energy probability contour is the 3 curve in Fig. \ref{fig:contour}. 
 A global analysis to combine all these informations is still in progress.

\section{Conclusions}
\label{sec:conclu}
 The MACRO results on atmospheric neutrinos were presented and discussed. 
  For all the categories of neutrino-induced muons detectable by MACRO, the 
observed zenith angle distributions and the numbers of events (Fig. 
\ref{fig:cos-uptr} and Fig. \ref{fig:low_cosze}) 
disagree with the predictions of the no oscillation hypothesis. The 
muon angular distribution and flux are in agreement with the hypothesis of 
two-flavour $\nmnt$ oscillations, with maximal mixing and
$\Dm2 = 2.5\cdot 10^{-3}$ eV$^2$. The hypothesis of $\nmns$ oscillations is 
disfavoured at the 99\% C.L. 

{
}
\end{document}